# SRF CAVITY GEOMETRY OPTIMIZATION FOR THE ILC WITH MINIMIZED SURFACE E.M. FIELDS AND SUPERIOR BANDWIDTH

N. Juntong, R.M. Jones and I.R.R. Shinton
The Cockcroft Institute, Daresbury, Warrington, Cheshire WA4 4AD, UK.
The University of Manchester, Oxford Road, Manchester, M13 9PL, UK.

*Abstract*

The main linacs of the ILC consist of nine-cell cavities based on the TESLA design. In order to facilitate reaching higher gradients we have re-designed the cavity shape. This leads to a reduction, comparable to several current designs, in both the ratio of the surface electric field to the accelerating field ($E_s/E_a$) and the magnetic field to the accelerating field ($B_s/E_a$). The bandwidth of the accelerating mode is also optimized. This new shape, which we refer to as the New Low Surface Field (NLSF) design, bears comparison with the Ichiro [1], Re-entrant [2] and LSF [3] designs.

## INTRODUCTION

The International Linear Collider (ILC) is a machine designed to produce a centre of mass energy of 500 GeV and with the potential to be upgraded to 1 TeV [4]. This lepton collider accelerates electrons and positrons through superconducting (SC) linear accelerators prior to colliding them at the interaction point (IP) of the machine. The present baseline design prescribes a working accelerating gradient of 31.5 MV/m. Increasing the accelerating gradient has the beneficial effect of potentially reducing the overall footprint of the accelerator. Increasing the accelerating gradient leads to increased surface electric and magnetic fields. An enhanced surface electric field ($E_s$) often gives rise to field emission of electrons and can lead to electron capture in the field of the r.f. accelerating cavities. On the other hand, increasing the surface magnetic field ($B_s$) can quench the superconducting properties of the cavity. Clearly it is advantageous to minimize both the surface magnetic field and the surface electric field with respect to the accelerating field ($E_a$). Another figure of merit is produced by the fractional operating bandwidth $k_c (\approx (\omega_\pi - \omega_0)/\omega_{\pi/2})$, of the monopole (accelerating) mode. This can be understood by considering the fractional frequency separation of the next nearest mode to the π accelerating mode:

$$\Delta \overline{\omega} = \frac{\omega_\pi - \omega_{(1-N^{-1})\pi}}{\omega_{\pi/2}} = \left(\frac{\pi}{2N}\right)^2 k_c \qquad (1)$$

where N = 9 for ILC cavities. Thus, in designing the geometry for a particular cavity shape we focused on optimizing three quantities $E_s/E_a$, $B_s/E_a$ and $k_c$; the former two are minimized and the later is maximized.

This paper is organized such that the procedure used to obtain the optimized parameters is described in the next section. The penultimate section analyses the higher order modes that will be excited in these NLSF cavities. The final section includes a summary and future work.

## OPTIMIZATION PROCEDURE

The standard SC cell geometry is illustrated in Fig. 1. The ellipticity of each surface is determined by the ratio of *a/b* and *A/B*. These surfaces are connected with a vertical surface as shown in Fig. 1. Several parameters are maintained at fixed values in order to make the optimization procedure practical. Regardless of optimization, the cell length is fixed at half the operating wavelength ($L = \lambda/2 = 115.304$ mm). We assign the iris thickness (*a*) to that of the LSF design and the equator radius ($R_{eq}$) to that of the LL [5, 6] design. All other parameters are not varied, other than the cavity parameters *b* and *B*. Thus, a two parameter optimization is performed with the goal to minimize $E_s/E_a$, $B_s/E_a$ and maximize $k_c$. We performed simulations with HFSS v11 [7] in the eigenmode module with a π phase advance per cell enforced. Throughout the simulations we ensured that, while the optimization progresses, the frequency is maintained at the correct accelerating frequency of 1.3 GHz. The frequency shift as a function of small changes in the elliptical parameter *B* is indicated in Fig. 2 for fixed values of *b*. This linear behavior allowed a rapid determination of the optimized parameter *B* at a prescribed *b*.

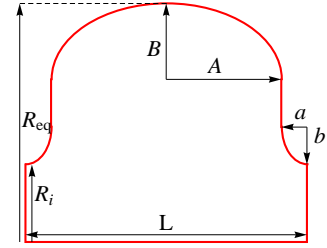

Figure 1: Cell parameters for a middle NLSF cell.

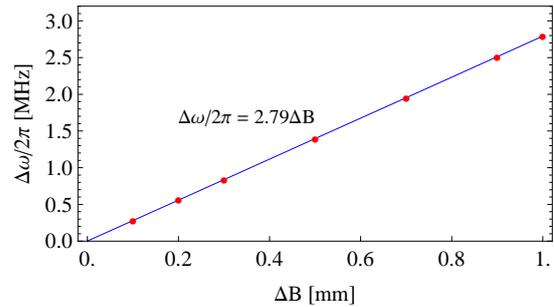

Figure 2: Frequency $\Delta\omega/2\pi$ versus elliptical parameter $\Delta B$.

Table 1: NLSF optimized parameters compared to TESLA, Ichiro and LL designs

| Parameter | TESLA | LL | ICHIRO | NLSF-1 | NLSF-2 | NLSF-3 | NLSF-4 | NLSF-5 | NLSF-6 | NLSF-7 | NLSF-8 | NLSF-9 |
|---|---|---|---|---|---|---|---|---|---|---|---|---|
| $R_i$ [mm] | 35 | 30 | 30 | 30 | 30 | 30 | 31 | 31 | 32 | 32 | 33 | 33 |
| $R_{eq}$ [mm] | 103.3 | 98.58 | 98.14 | 98.14 | 98.58 | 100 | 98.14 | 98.58 | 98.14 | 98.58 | 98.14 | 98.58 |
| A [mm] | 42 | 50.052 | 50.052 | 47.152 | 47.152 | 47.152 | 47.152 | 47.152 | 47.152 | 47.152 | 47.152 | 47.152 |
| B [mm] | 42 | 36.5 | 34.222 | 30.35 | 33 | 41.45 | 30.1 | 32.7 | 28.65 | 31.35 | 27.2 | 29.8 |
| $a$ [mm] | 12 | 7.6 | 7.6 | 10.5 | 10.5 | 10.5 | 10.5 | 10.5 | 10.5 | 10.5 | 10.5 | 10.5 |
| $b$ [mm] | 19 | 10 | 9.945 | 20.15 | 19 | 17.2 | 16.5 | 15.5 | 16.5 | 15.5 | 16.5 | 15.5 |
| $f_s$ [GHz] | 1.30116 | 1.30008 | 1.30040 | 1.30000 | 1.29998 | 1.30002 | 1.30012 | 1.30000 | 1.30002 | 1.30023 | 1.30006 | 1.30003 |
| Bandwidth [MHz] | 24.32 | 19.74 | 19.82 | 17.25 | 16.92 | 16.23 | 18.52 | 18.17 | 20.86 | 20.50 | 23.38 | 22.98 |
| $k_c$ [%] | 1.89 | 1.53 | 1.54 | 1.34 | 1.31 | 1.26 | 1.43 | 1.41 | 1.62 | 1.59 | 1.81 | 1.78 |
| $E_s/E_a$ | 2.18 | 2.42 | 2.37 | 2.07 | 2.06 | 2.07 | 2.09 | 2.12 | 2.12 | 2.11 | 2.16 | 2.18 |
| $B_s/E_a$ [mT/(MV/m)] | 4.18 | 3.64 | 3.62 | 3.78 | 3.75 | 3.76 | 3.81 | 3.79 | 3.86 | 3.83 | 3.91 | 3.88 |

The behavior of the three figures of merit $E_s/E_a$, $B_s/E_a$ and $k_c$ as a function of iris radius are illustrated in Fig. 3. For example a 1 mm increase in iris radius results in an increase of $k_c$ by ~12-13%, surface electric field ratio by 2% and surface magnetic field by 1.3%. Several designs were considered during this optimization process and a selection are illustrated in Table 1. The final value chosen is highlighted. This design has a 8.5% lower $B_s/E_a$ field compared to TESLA, and $E_s/E_a$ is 13% lower than the LL design. Both ratios $E_s/E_a$ and $B_s/E_a$ are comparable to the LSF design; however the bandwidth is superior as it is 26.5% wider.

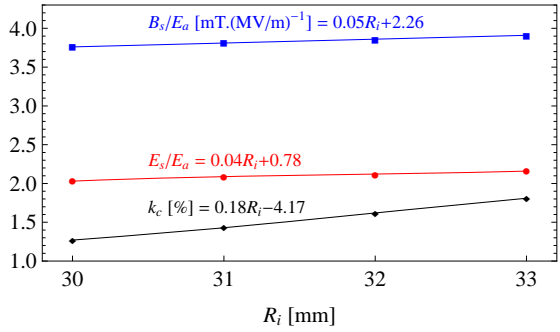

Figure 3: Three figures of merit, $k_c$, $E_s/E_a$ and $B_s/E_a$ versus iris radius

The salient figures of merit for a range of potential new designs compared to current designs are displayed in Fig. 4. The corresponding surface e.m. fields are displayed in

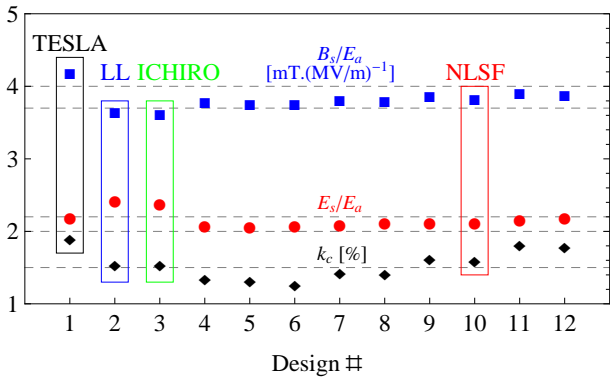

Figure 4: Three figures of merit of NLSF compared to other designs.

Fig. 5 together with the final optimized geometry in Fig. 6. In all cases, the electric field reaches a peak value close to the tip of the iris. Similarly, the magnetic field is slightly intensified for NLSF and LL cavities close to the iris tip. The enhanced bandwidth in the NLSF geometry facilitates less sensitivity to fabrication errors as the monopole mode separation is larger than other current designs.

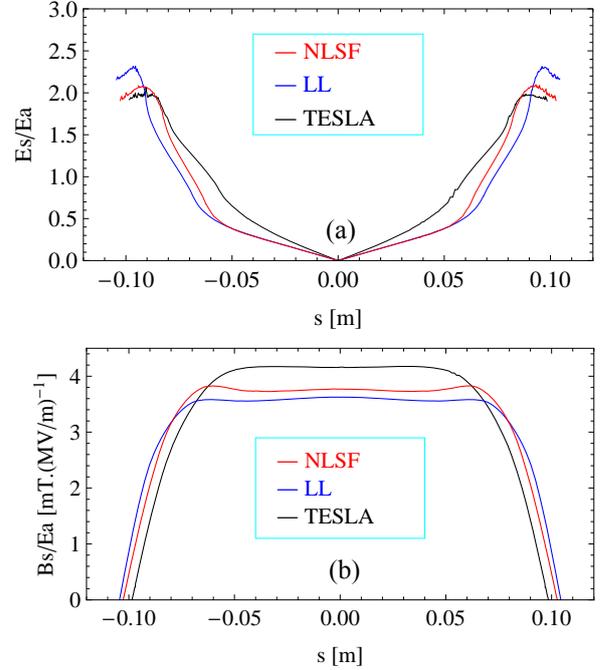

Figure 5: Surface electric field (a) and magnetic field (b) compared to TESLA (black), LL (blue) and NLSF (red) designs. Here $s$ is the coordinate along the surface of the cavity.

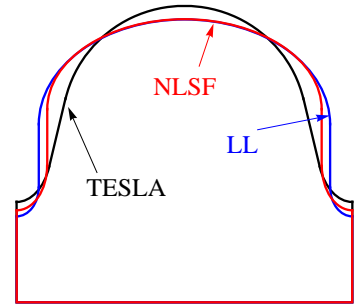

Figure 6: NLSF optimized cavity geometry compared to alternative designs.

The beam also excites a wakefield which can be analyzed in terms of a series of cavity modes. Provided these are well-damped the quality of the beam is maintained. These higher order modes are discussed in the next section.

## NLSF HIGHER ORDER MODES

The ultra-relativistic beam excites a wakefield which has both longitudinal and transverse components. The former can give rise to an energy spread along the bunches and the later can, at minimum dilute the emittance of the beam or, in the worst case, can cause a BBU [8] instability. The wakefield influences particles within a single bunch (the short range or intra-bunch) wakefield and the multi-bunch (the long range or inter-bunch) wakefield which effects successive bunches in a train of accelerated bunches. Here we consider the transverse component of the long-range wakefield. The transverse wakefield can be decomposed into the modes excited in the SC cavities. We study the dipole modes excited in the NLSF cavity. The brillouin diagram for the first 6 dipole bands is illustrated in Fig. 7 together with neighbouring sextupole bands. The curves were obtained from simulations entailing a quarter cell subjected to symmetry conditions appropriate to dipole mode excitation (electric wall is perpendicular to the magnetic wall). The points correspond to the modes in a 9-cell structure. Provided the frequencies are below the end iris's terminating cut-off frequency, the eigenmodes are relatively independent of the terminating planes at either end of the cavity. In order to assess the impact of these modes on the momentum kick to the beam we also calculated the R/Q factors and these are illustrated for the dipole modes in Fig. 8, simulated with HFSS v11 up to the 6$^{th}$ dipole band.

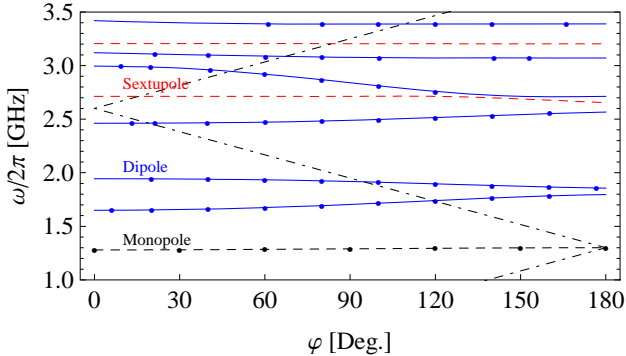

Figure 7: Brillouin diagram of a NLSF cavity. Cells subjected to an infinite periodic structure are indicated by the lines. Monopole, dipole and sextupole modes are indicated by the following lines: black dashed, solid blue and red dashed respectively. The points indicate cavity eigenfrequencies.

The modes with large R/Q values are given by those synchronous with the beam and they correspond to the region close to the intersection of the light line with the dipole bands in Fig. 8. For example the largest R/Qs for each of the first three bands are located at phase advances of $2\pi/3$, $5\pi/9$ and $\pi/9$, with the largest being located in the 3$^{rd}$ dipole band. The LSF structure has a similar mode structure [3].

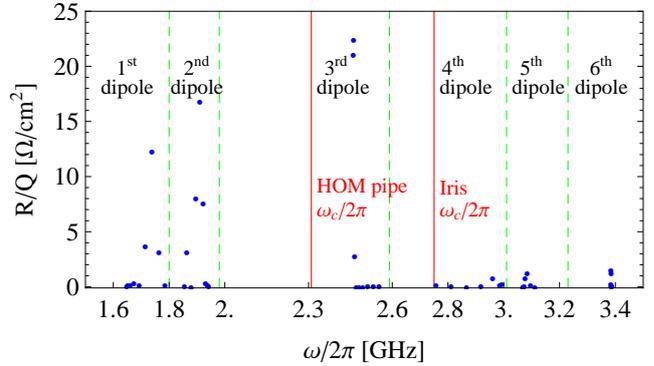

Figure 8: Dipole mode R/Qs for a nine-cell NLSF cavity. The cut-off frequencies of the HOM pipe and iris are indicated by $\omega_c/2\pi$.

## SUMMARY

A new design, based on the LL and LSF cavities allows parameter optimization, a reduction in the ratio of both $B_s/E_a$, $E_s/E_a$ and an increased monopole mode fractional bandwidth $k_c$. This NLSF design has an 8.5% reduced $B_s/E_a$ ratio compared to TESLA shape and a 13% reduced $E_s/E_a$ compared to the LL design. In addition the bandwidth is 26.5% wider and $B_s/E_a$ together with $E_s/E_a$ are comparable to the LSF design. The higher order modes, affecting multiple bunches within the train, indicate that, as is usual in these L band cavities, the first three dipole bands have large R/Q values. Suitable couplers must be designed in order to adequately damp these modes. Further work [9], will entail detailed studies of the monopole and dipole modes in NLSF cavities equipped with redesigned end cells and HOM couplers.

## ACKNOWLEDGEMENT

We have benefited from discussions at the weekly Manchester Electrodynamics and Wakefield (MEW) meetings held at the Cockcroft Institute, where these results were first presented. N.J. is in receipt of joint support from the Royal Thai Government and the Thai Synchrotron Light Research Institute.

## REFERENCES

[1] T. Saeki et al., EPAC06, MOPLS087, 2006.
[2] V. Shemelin, PAC05, TPPT068, 2005.
[3] Z. Li, C. Adolphsen, LINAC08, THP038, 2008.
[4] The International Linear Collider Reference Design Report, 2007.
[5] J. Sekutowicz et al., JLAB TN-02-023, 2002.
[6] J. Sekutowicz et al., PAC05, TPPT056, 2005.
[7] www.ansoft.com/products/hf/hfss/
[8] K. Yokoya, DESY Report 86-084, 1986.
[9] To be submitted to SRF2009.